\documentclass[twocolumn,prb]{revtex4-1}

\usepackage{graphicx}
\usepackage{epsfig}
\usepackage{amsmath, amssymb}
\usepackage{verbatim}

\begin{document}

\title{Phase transitions in $Z_N$ gauge theory and twisted $Z_N$ topological phases}

\author{Maissam Barkeshli}
\affiliation{Department of Physics, Stanford University, Stanford, CA  94305}
\author{Xiao-Gang Wen}
\affiliation{Department of Physics, Massachusetts Institute of Technology,
Cambridge, MA 02139 }

\begin{abstract}

We find a series of non-Abelian topological phases that are 
separated from the deconfined phase of $Z_N$ gauge theory by a 
continuous quantum phase transition. These non-Abelian states, 
which we refer to as the ``twisted'' $Z_N$ states, are described 
by a recently studied $U(1) \times U(1) \rtimes Z_2$ Chern-Simons 
(CS) field theory. The $U(1) \times U(1) \rtimes Z_2$ CS theory 
provides a way of gauging the global $Z_2$ electric-magnetic 
symmetry of the Abelian $Z_N$ phases, yielding the twisted $Z_N$ states. 
We introduce a parton construction to describe the Abelian $Z_N$ phases 
in terms of integer quantum Hall states, which then allows us to 
obtain the non-Abelian states from a theory of $Z_2$ fractionalization. 
The non-Abelian twisted $Z_N$ states do not have topologically protected gapless edge modes 
and, for $N > 2$, break time-reversal symmetry. 

\end{abstract}

\maketitle

\section{Introduction}

Landau symmetry breaking theory\cite{L3726,LanL58} not only classifies a large
class of symmetry breaking phases, it also tell us which pairings of phases
are connected by continuous phase transitions.  Now we know that there are many
topologically ordered states\cite{Wrig} that cannot be described
by Landau symmetry breaking theory.  The next important issue is to understand
which pairs of topological phases are connected by continuous phase
transitions and what the critical properties of the transitions are. 

One class of topological phases are Abelian fractional quantum Hall (FQH)
states. Those states can be systematically described by the $K$-matrix and the
associated $U(1)\times U(1)\times...$ Chern-Simons (CS)
theory.\cite{BW9045,R9002}  For such a class of topological
states, we know that two Abelian FQH states described by $K$ and 
\begin{align}
K' =\bpm K & l\\
         l^T & p\\
\epm  ,
\end{align}
are connected by continuous phase transitions, provided that there is a periodic
potential with a proper period.\cite{WW9301,CFW9349}  This class of phase
transitions is induced by anyon condensation described by Ginzburg-Landau
Chern-Simons theory.\cite{ZHK8982} Another class of topological phases are
described by gauge theory with a discrete, possibly non-Abelian, gauge
group $G$.  For such classes of non-Abelian topological
states,\cite{LP9321,BD9263} we also know that a pair of discrete gauge theories
described by gauge groups $G$ and $G'$ can be connected by continuous phase
transitions if $G' \subseteq G$ or $G \subseteq G'$.  This class of phase transitions are
induced by boson condensation described by the standard Anderson-Higgs
mechanism of ``gauge symmetry'' breaking.\cite{FS7982,BS0916}

But there exist more general classes of topological phases described by pattern
of zeros,\cite{WW0808,WW0809,BW0932,BW1001a} $Z_n$ vertex
algera,\cite{MR9162,WW9455,LW1024} and/or string-net
condensates.\cite{LWstrnet}  The above picture of topological phase transitions
is clearly incomplete.  One attempt to find new classes of continuous
topological phase transitions is introduced in \Ref{BW1004,BW1070}, which
describe a class of continuous phase transitions between Abelian FQH and
non-Abelian FQH\cite{MR9162,Wnab} states, induced by anyon condensation. A
special case of such class of continuous topological phase transition is
induced by fermion condensation, which was first discussed in
\Ref{SMF9945,RG0067,W0050}

In \Ref{BW1023}, we studied $U(1) \times U(1) \rtimes Z_2$ CS theory with
integral coupling constants $(k,l)$ (see eq. \ref{CSkl}).  When $l = 3$, it was
found that the topological properties of this theory agree with those of the
$Z_4$ parafermion FQH states\cite{RR9984} at filling fraction $\nu = k/(2k-3)$, leading us
to suggest that this was the long wavelength field theoretic description of
these non-Abelian FQH states.  This formulation of the effective field theory
allowed us to show that there is a continuous phase transition, in the 3D Ising
universality class, between the $Z_4$ parafermion states and the Abelian
$(k,k,k-3)$ states in bilayer quantum Hall systems.\cite{BW1004} 

Subsequently, it was found that for more general values of the coupling
constants, $k,l \neq 0$, the $U(1) \times U(1) \rtimes Z_2$ CS theory describes
a series of non-Abelian FQH states -- the orbifold FQH states.\cite{BW1070} The
$Z_4$ parafermion FQH states are then a special case of these more general
orbifold FQH states, which are separated from the $(k,k,k-l)$ states by a
continuous 3D Ising phase transition. 

However, it is also known that $U(1) \times U(1)$ CS theory need not describe
only quantum Hall states.  For a particular set of coupling constants, 
it can also describe time-reversal invariant topological phases. The Lagrangian
\begin{align}
\label{mCS}
\mathcal{L} = \frac{1}{4\pi}\sum_{IJ} K_{IJ} a_I \partial a_J,
\end{align}
with $K = \left( \begin{matrix} 0 & N \\ N & 0 \end{matrix} \right)$, 
describes the long-wavelength properties of the deconfined phase of 
$Z_N$ gauge theory. Such a phase has $N^2$ topologically distinct, Abelian
quasiparticles. The quasiparticles can be labelled by their electric
and magnetic charge, $(e,m)$, for $e,m = 0, ..., N-1$, and have spin $em/N$. 

The above observation suggests that the $U(1) \times U(1) \rtimes Z_2$ CS
theory, with $k = 0$, may also describe non-Abelian topological phases,
although ones that may exist in frustrated spin models and that do not require
the presence of a strong external magnetic field. This raises many questions
regarding whether such phases can be obtained from microscopic models with
local interactions, how to develop a more complete theory for these possible
states, how to understand the conditions under which they may occur, and how to
understand their full topological order.

In this paper, we study the topological properties of these non-Abelian states
and, by developing a slave-particle description of them, show that they are
physical in that they can be realized in a microscopic model with local
interactions. This leads to a series of non-Abelian states, without protected
gapless edge modes, that are separated from the Abelian $Z_N$ states by a continuous 3D
Ising phase transition. For $N > 2$, these non-Abelian states can only be
accessed when time-reversal symmetry is broken -- though unlike
quantum Hall states, their existence does not require a large external
magnetic field whose flux is proportional to the number of particles.

The specific system that we use to establish our results is one with two
flavors of strongly interacting bosons. However, the main purpose of
the present work is to show that the topological phases and phase
transitions that we discuss here exist in principle and to understand their topological
properties; they may also appear in other
models with different microscopic degrees of freedom. 

We begin in Section \ref{ZNparton} by developing a parton construction for the Abelian $Z_N$
states, which allows us to describe these conventional phases in
terms of integer quantum Hall states. The use of this formulation is
that it allows us to access the $U(1) \times U(1) \rtimes Z_2$ phases
through a theory of $Z_2$ fractionalization. In Section \ref{slaveIsingSect}, we then
use this parton construction, in conjunction with a recently developed
slave Ising formulation, to describe the twisted $Z_N$ phases
and argue that their low energy theory should be the $U(1) \times
U(1) \rtimes Z_2$ CS theory with suitably chosen coupling constants. 
In Section \ref{Z2CS}, we review the results of the $U(1) \times U(1) \rtimes Z_2$
CS theory. In Section \ref{cftCon}, we give a prescription, using conformal field theory,
to derive the full topological order of these non-Abelian states; wherever
comparison is possible, we find agreement with highly non-trivial results 
from the $U(1) \times U(1) \rtimes Z_2$ CS theory.

\section{Parton construction for $Z_N$ topological order }
\label{ZNparton}

In this section we show how to construct a state with $Z_N$ topological
order by projecting from $\nu = 1$ IQH states\cite{J9153}. 

We begin with two flavors of bosons, $b_\uparrow$ and $b_\downarrow$, and decompose them
in terms of $3N$ partons as follows:
\begin{align}
\label{partonConsZN}
b_\uparrow &= \prod_{i = 1}^N \psi_i \prod_{j = 2N+1}^{3N} \psi_j,
\nonumber \\
b_\downarrow &= \prod_{i = N+1}^{2N} \psi_i \prod_{j = 2N+1}^{3N} \psi_j,
\end{align}
where we have suppressed the space indices. 
Note that the partons $\psi_{2N+1}, ..., \psi_{3N}$ are shared between $b_\uparrow$ and $b_\downarrow$. 
We can rewrite the original theory of these two flavors of bosons
in terms of a theory of these partons coupled to a gauge field. The gauge field
projects the expanded parton Hilbert space onto the physical Hilbert space,
which is generated by the physical operators $b_\uparrow$ and $b_\downarrow$. 

Next, we assume a mean-field ansatz where $\psi_1,...,\psi_{2N}$ 
form a $\nu = 1$ IQH state, while $\psi_{2N+1},...,\psi_{3N}$ form 
a $\nu = -1$ IQH state. The maximal gauge group that respects this
mean-field ansatz is $SU(N) \times SU(N) \times SU(N) \times U(1)$,
which we will write as $SU(N)^3 \times U(1)$.

In order to motivate the above construction, note that bilayer $(NN0)$
FQH states can be obtained through the parton construction by decomposing
the electron operator in each layer as
\begin{align}
\Psi_{e\uparrow} &= \psi_1 \cdots \psi_N,
\nonumber \\
\Psi_{e\downarrow} &= \psi_{N+1} \cdots \psi_{2N},
\end{align}
and assuming a mean-field ansatz where the $\psi_i$ each form $\nu = 1$ IQH
states. It can be shown that the low energy field theory for such a state is $U(1) \times U(1)$
CS theory with $K$-matrix $K = \left( \begin{matrix} N & 0 \\ 0 & N \end{matrix} \right)$.
In order to describe more general bilayer FQH states such as $(N+m, N+m, m)$, we simply multiply
each electron operator by an additional set of operators:
\begin{align}
\label{partDecZN1}
\Psi_{e\uparrow} &= \psi_1 \cdots \psi_N \times \psi_{2N+1} \cdots \psi_{2N+m} ,
\nonumber \\
\Psi_{e\downarrow} &= \psi_{N+1} \cdots \psi_{2N} \times \psi_{2N+1} \cdots \psi_{2N+m},
\end{align}
and we assume again that all of the partons $\psi_i$ form a $\nu = 1$ IQH state.
At the level of the wave functions, this has the effect of multiplying the
$(NN0)$ wave function by a Jastrow factor to give the $(N+m, N+m, m)$
wave function:
\begin{align}
\label{wfnZN1}
\Phi_{(N+m,N+m,m)} = \Phi_{(NN0)} \Phi_{(mmm)}.
\end{align}
Since the $Z_N$ gauge theory is described by a $K$-matrix 
\begin{align}
\label{KZN1}
K = \left( \begin{matrix} 0 & N \\ N & 0 \end{matrix} \right) = 
\left( \begin{matrix} N & N \\ N & N \end{matrix} \right) - 
\left( \begin{matrix} N & 0 \\ 0 & N \end{matrix} \right) ,
\end{align}
a natural guess is the decomposition in (\ref{partonConsZN}), where 
$\psi_{2N+1}, \cdots, \psi_{3N}$ are assumed to form $\nu = -1$ IQH states. 

The low energy theory for such a state will involve the partons interacting
with a gauge field from the gauge group $SU(N)^3 \times U(1)$. 
It is not at all clear that such a complicated field theory, with many non-Abelian
gauge groups, has simply the $Z_N$ topological order.

In Appendix \ref{NAtorusCalc}, we compute the ground state degeneracy of the 
$SU(N)^3 \times U(1)$ theory on a torus.
We find that it is given by
\begin{align}
\text{Torus Degeneracy} = N^2,
\end{align}
which agrees with that of the $Z_N$ topological order. 

Unfortunately, besides the torus ground state degeneracy, it is 
extremely difficult to compute any other topological
properties of a theory with such a complicated non-Abelian gauge group. 
In order to proceed, we choose a mean-field ansatz for the partons 
that breaks the gauge group down to the center of 
$SU(N)^3 \times U(1)$, which is $U(1)^{3N-2}$. One way to do this,
for example, is to assume various condensates such that in the low 
energy field theory, the partons all have different masses, while
still forming the IQH states described above. Since the gauge group 
is now Abelian, it is possible to compute all topological
properties of the resulting states. In the following section, we show that 
such a gauge theory coupled to the partons describes the topological 
properties of the $Z_N$ phase and directly yields the $U(1) \times U(1)$ mutual 
CS theory as its low energy effective field theory.

\subsection*{Mutual $U(1) \times U(1)$ CS theory from parton construction}

The effective field theory is described by the Lagrangian:
\begin{equation}
\label{partonLDZN}
\mathcal{L} = i \psi^{\dagger} \partial_0 \psi +  \psi^{\dagger} \frac{M^{-1}}{2} (\partial - i A Q)^2 \psi 
+ \text{Tr }(j^\mu a_\mu^I p^I) + \cdots,
\end{equation}
where $\psi^T = (\psi_1, \cdots, \psi_{3N})$,
$M_{ab} = m_a \delta_{ab}$ and $m_a$ is the mass of the $a$th parton,
$A_i$ describes a magnetic field seen by the partons, $j^\mu_{ab} = \psi_a \partial^\mu \psi_b$ 
describes the current of the partons, $a_\mu$ is the gauge field, and
the $3N-2$ generators of the gauge group 
$U(1)^{N-1} \times U(1)^{N-1} \times U(1)^{N-1} \times U(1)$
are given by the matrices $p^I$:
\begin{align}
p^I_{ij} &= \delta_{ij} (\delta_{i,I} - \delta_{i, I + 1}), \;\;\;\;\; I = 1, \cdots, N-1,
\nonumber \\
p^I_{ij} &= \delta_{ij} (\delta_{i,I+1} - \delta_{i,I+2}), \;\;\;\;\; I = N, \cdots, 2N-2,
\nonumber \\
p^I_{ij} &= \delta_{ij} (\delta_{i,I+2} - \delta_{i,I+3}), \;\;\;\;\; I = 2N-1, \cdots, 3N-3,
\nonumber \\
p^I_{ij} &= \delta_{ij} (\delta_{i,1} + \delta_{i,N+1} - \delta_{i, 2N +1}), \;\;\;\;I = 3N-2.
\end{align}

Since the partons are in $\nu = 1$ IQH states, their action is each given by a
$U(1)_1$ CS theory; because of the gauge constraint they will be coupled to the gauge field as well:
\begin{align}
\label{dualCSZN}
&\mathcal{L} = \mathcal{L}_{parton} + \mathcal{L}_{constraint}
\nonumber \\
&\mathcal{L}_{parton} = \frac{1}{4\pi} \sum_{i = 1}^{2N} b^i \partial b^i - \frac{1}{4\pi} \sum_{i=2N+1}^{3N} b^i \partial b^i
\nonumber \\
&\mathcal{L}_{constraint} = j^i_\mu p_{ij}^I \delta_{ij} a_\mu^I,
\end{align}
where $b^i$ is a $U(1)$ gauge field describing the current density of the $i$th parton:
\begin{align}
j^i_\mu = \frac{1}{2\pi} b^i_\mu \partial_\nu b^i_\lambda.
\end{align}

From the definition of the $p^I$, we see that integrating out the $a$
gauge fields enforces the constraints:
\begin{align}
j^1 = \cdots = j^N, \;\;\;\; & j^{N+1} = \cdots = j^{2N}, 
\nonumber \\
j^{2N+1} = \cdots = j^{3N}, \;\;\;\; & j^{2N+1} = j^{N+1} + j^1.
\end{align}
Therefore, the effective action becomes:
\begin{align}
\label{dualAction}
\mathcal{L} = -\frac{N}{4\pi} (b^1 \partial b^{N+1} + b^{N+1} \partial b^1),
\end{align}
which is exactly the action for the mutual $U(1) \times U(1)$ CS theory description
of $Z_N$ topological order. Actually this analysis is essentially the same analysis 
that we intuited by analyzing wave functions in (\ref{partDecZN1}) - (\ref{KZN1}).

When the masses of the partons are all equal, we see that the theory has the
enhanced $SU(N) \times SU(N) \times SU(N) \times U(1)$ gauge symmetry.  Since
the number of states on the torus does not change when this gauge symmetry is
broken to its Abelian subgroup by assuming different mean-field masses for the
partons, we conjecture that it also describes the topological properties of the
$Z_N$ phases. This is a surprising result, for it provides an example in which
gauge symmetry breaking does not actually change the topological properties of
a state.  This is related to the fact that sometimes a non-Abelian CS theory
is equivalent to an Abelian CS theory. For example non-Abelian $SU(k)$ level 1
CS theory is equivalent to Abelian $U(1)$ level $k$ CS theory.\cite{W8951,FMCFT}

\section{Slave Ising description}
\label{slaveIsingSect}

The parton description presented in the previous section
yields the mutual $U(1) \times U(1)$ CS theory at long wavelengths
in a way that is amenable to a certain $Z_2$ ``twisting.''

To do this, we follow the slave-Ising construction presented
in \Ref{BW1070} in the context of the orbifold non-Abelian FQH states. 
We start with two boson operators defined on a lattice, 
$b_{i\sigma}$, and we consider the positive and negative combinations:
\begin{align}
\label{PsipmZN}
b_{i\pm} &= \frac{1}{\sqrt{2}} (b_{i\uparrow} \pm b_{i\downarrow}).
\end{align}
We introduce two new fields at each lattice site $i$: 
an Ising field $s^z_i = \pm 1$ and a bosonic field $d_{i-}$, 
and we rewrite $b_{i-}$ as
\begin{align}
b_{i+} \equiv d_{i+}, \;\;\;\;
b_{i-} = s^z_i d_{i-}.
\end{align}
This introduces a local $Z_2$ gauge symmetry, associated with the transformation
\begin{align}
s^z_i \rightarrow -s^z_i, \;\;\;\; d_{i-} \rightarrow -d_{i-}.
\end{align}
The electron operators are neutral under this $Z_2$ gauge symmetry, and therefore
the physical Hilbert space at each site is the gauge-invariant set of states at
each site:
\begin{align}
(|\uparrow \rangle + |\downarrow \rangle) &\otimes |n_{d_{-}} = 0 \rangle
\nonumber \\
(|\uparrow \rangle - |\downarrow \rangle) &\otimes |n_{d_{-}} = 1 \rangle,
\end{align}
where $|\uparrow \rangle$ ($|\downarrow \rangle$) is the state with
$s^z = +1 (-1)$, respectively. In other words, the physical states at each 
site are those which satisfy
\begin{align}
\label{IsingConstraintZN}
(s^x_i + 1)/2 + n_{d_{i-}}  = 1.
\end{align}

If we imagine that the bosons $d_{i \pm}$ form some gapped state, then we 
would generally expect two distinct phases \citep{SF0050}: the deconfined/$Z_2$ unbroken
phase, where
\begin{align}
\langle s^z_i \rangle = 0,
\end{align}
and the confined/Higgs phase, where upon fixing a gauge we have
\begin{align}
\langle s^z_i \rangle \neq 0.
\end{align}
We seek a mean-field theory where the deconfined phase has the
properties described by the $U(1) \times U(1) \rtimes Z_2$ CS theory, 
and the confined/Higgs phase corresponds to the $Z_N$ topological
phases. To do this, observe that in the Higgs phase we have
\begin{align}
b_{i \pm} = d_{i \pm}, 
\end{align}
since we may set $s^z_i = 1$ in this phase. Now for this to describe the 
$Z_N$ phases, we use the parton construction of Section \ref{ZNparton}:
\begin{align}
d_{i\pm} &= \frac{1}{\sqrt{2}} (d_{i1} \pm d_{i2}),
\nonumber \\
d_{i1} &= \psi_{1i} \cdots \psi_{Ni} \psi_{2N+1,i} \cdots \psi_{3N,i} ,
\nonumber \\
d_{i2} &= \psi_{N+1,i} \cdots \psi_{2N,i} \psi_{2N+1,i} \cdots \psi_{3N,i} ,
\end{align}
and we assume that $\psi_1,..., \psi_{2N}$ form a $\nu = 1$ IQH state while
$\psi_{2N+1},..., \psi_{3N}$ form a $\nu = -1$ IQH state. 

Clearly, the low energy field theory of the confined phase is the 
mutual $U(1) \times U(1)$ CS theory, describing the Abelian $Z_N$
topological order. In the deconfined phase, we see that the parton sector 
is still described by a $U(1) \times U(1)$ CS theory, but that there is
also an additional $Z_2$ gauge symmetry associated with exchanging
the two $U(1)$ gauge fields. This is precisely the content
of the $U(1) \times U(1) \rtimes Z_2$ CS theory,\cite{BW1023} which we therefore 
expect to describe the topological properties of this $Z_2$ deconfined phase. 

Since the transition between these two phases is induced by the condensation
of the Ising spin $s^z_i$, which is coupled to a $Z_2$ gauge field, we see that
as the gap to the $s^z$ excitations is reduced, the low energy field theory is
simply a real scalar field coupled to a $Z_2$ gauge field. Such a theory
was analyzed in \Ref{FS7982}, where it was found that the transition is continuous and
in the 3D Ising universality class. Therefore, the Abelian $Z_N$ and its
$Z_2$ fractionalized neighbor, the ``twisted'' $Z_N$ states, are separated by
a continuous quantum phase transition. 

A useful property of this slave Ising formulation is that standard methods
of constructing projected trial wave functions will, when applied to the
$Z_2$ deconfined phase, yield possible trial wave functions for these non-Abelian
twisted $Z_N$ states.\cite{Wen04,BW1070} 

\section{$U(1) \times U(1) \rtimes Z_2$ CS theory and topological quantum numbers of
twisted $Z_N$ states}
\label{Z2CS}

 
The $U(1) \times U(1) \rtimes Z_2$ CS theory was studied in detail in \Ref{BW1023}.
In this section, we review the results for the choice of
coupling constants that is relevant here. 

The $U(1) \times U(1) \rtimes Z_2$ CS theory is described by the Lagrangian
\begin{align}
\label{CSkl}
\mathcal{L} = \frac{k}{4\pi} (a \partial a + \tilde{a} \partial \tilde{a})
+ \frac{k-l}{4\pi} (a \partial \tilde{a} + \tilde{a} \partial a),
\end{align}
where $a$ and $\tilde{a}$ are two $U(1)$ gauge fields.  Formally, this is the
same Lagrangian as that of the $U(1) \times U(1)$ CS theories, although here we
also have an additional $Z_2$ gauge symmetry associated with interchanging the
two $U(1)$ gauge fields at each space-time point. This allows, e.g., for the
possibility of $Z_2$ vortices -- configurations in which the two $U(1)$ gauge
fields transform into each other around the vortex -- and twisted sectors on
manifolds of non-trivial topology.

In order to describe the twisted $Z_N$ topological phases, we choose $k = 0$
and $l = N$. In \Ref{BW1023}, we found that such a theory has $N (N + 7)/2$ topologically 
distinct quasiparticles. The ground state degeneracy on genus $g$ surfaces is
\begin{align}
S_g(N) = (N^g/2) [N^g + 1 + (2^{2g} - 1)(N^{g-1} + 1) ] .
\end{align}
From $S_g(N)$ we can obtain the quantum dimensions of all the quasiparticles. 
The total quantum dimension is
\begin{align}
D^2 = 4N^2.
\end{align}
There are three classes of quasiparticles: $2N$ quasiparticles with quantum dimension 
$d = 1$, $2N$ quasiparticles with quantum dimension $d = \sqrt{N}$, and 
$N(N-1)/2$ quasiparticles with quantum dimension $d = 2$. 

The fundamental non-Abelian excitations in the $U(1) \times U(1) \rtimes Z_2$
CS theory are $Z_2$ vortices. In \Ref{BW1023}, we studied the number of degenerate
ground states in the presence of $n$ pairs of $Z_2$ vortices at fixed
locations on a sphere. The result for the number of such states is:
\begin{equation}
\alpha_n =  \left\{
\begin{array}{ll}
    (N^{n-1} + 2^{n-1})/2 &\mbox{ for $N$ even,} \\
    (N^{n-1} + 1)/2 & \mbox{ for $N$ odd.} \\
    \end{array} \right.
\end{equation}
This shows that the quantum dimension of the $Z_2$ vortices is
$d = \sqrt{N}$. We can also compute the number of states that are
odd under the $Z_2$ gauge transformation. The number of these $Z_2$ 
non-invariant states turns out to be an important quantity, because
it yields important information about the fusion rules of the quasiparticles.
The number of $Z_2$ non-invariant states yields the number of ways
for $n$ pairs of $Z_2$ vortices to fuse to an Abelian quasiparticle
that carries $Z_2$ gauge charge. The ground state degeneracy of 
$Z_2$ non-invariant states in the presence of $n$ pairs of
$Z_2$ vortices at fixed locations on a sphere was computed to be
\begin{equation}
\beta_n = \left\{ 
\begin{array}{ll}
    (N^{n-1} - 2^{n-1})/2 &\mbox{ for $N$ even,} \\
    (N^{n-1} - 1)/2 & \mbox{ for $N$ odd.} \\
    \end{array} \right. 
\end{equation}
Thus if $\gamma$ labels a $Z_2$ vortex, these calculations reveal
the following fusion rules for $\gamma$ and its conjugate $\bar{\gamma}$:
\begin{align}
(\gamma \times \bar{\gamma})^n = \alpha_n \mathbb{I} + \beta_n j + \cdots,
\end{align}
where $j$ is a topologically non-trivial excitation that carries the $Z_2$
gauge charge. The $\cdots$ represent additional quasiparticles that
may appear in the fusion. 

Note that the above is true also for $U(1) \times U(1) \rtimes Z_2$ CS theory
with coupling constants $(k,l) = (N, 0)$, which applies to bilayer FQH states. 
This indicates a close relation between the FQH phases with $(k,l) = (N, 0)$ 
and the non-quantum Hall ones with $(k,l) = (0, N)$

The above gives us much information about the topological order of the
twisted $Z_N$ states, but we have not been able to compute the full
topological order of these states directly form the $U(1) \times U(1) \rtimes Z_2$
CS theory. However, since we know that the twisted $Z_N$ states contain
a $Z_2$ charged boson -- labelled $s^z_i$ in the previous section and $j$ here --
whose condensation yields the Abelian $Z_N$ states, we can deduce even more
topological properties of the quasiparticles. 

In our case, the two phases are separated by the condensation of a topologically
non-trivial bosonic quasiparticle, $j$, that fuses with itself
to a local topologically trivial excitation. Based on general considerations,\cite{BS0916} 
we expect the following regarding the topological quantum numbers of such phases.
Upon condensation of $j$, quasiparticles that differed from each other by fusion with
$j$ become topologically equivalent. Quasiparticles that were non-local with respect
to $j$ before condensation become confined after condensation and do not appear in
the low energy spectrum. Finally, quasiparticles that fused with their conjugate to the
identity and $j$ will, after condensation, split into two topologically distinct quasiparticles.
The spins of the quasiparticles remain unchanged through this process, which allows us
to obtain information about the spins of some of the quasiparticles in the 
twisted $Z_N$ states from knowledge of the spins of the quasiparticles in the 
Abelian $Z_N$ states.

In the case of the twisted $Z_N$ states, we have the following.
The $2N$ Abelian quasiparticles, which contain the quasiparticle $j$, become
$N$ Abelian quasiparticles after condensation. The $Z_2$ vortices are clearly non-local
with respect to the $Z_2$ charges, so they become confined. Finally, the $N(N-1)/2$
quasiparticles with quantum dimension 2 each split into two distinct quasiparticles.
This yields the $N^2$ quasiparticles of the Abelian $Z_N$ states. The natural 
interpretation is that the $N(N-1)/2$ quasiparticles correspond to the $Z_2$ invariant
combinations of quasiparticles in the Abelian states: $(e,m) + (m,e)$ for $e \neq m$, 
while the $2N$ Abelian quasiparticles of the twisted $Z_N$ states consist of the $N$ diagonal
quasiparticles $(l,l)$, and their $N$ counterparts that differ by fusion with $j$. 
Therefore we can infer the spins of these two classes of quasiparticles. The results
are listed in Table \ref{qpContent}. 

\begin{table}
\begin{tabular}{ccc}
 & Spin & Quantum Dimension \\
\hline
$A_l$  & \;\;\;\;\; $l^2/N$ \;\;\;\;\; & 1 \\
\\
$B_l$   & -  & $\sqrt{N}$  \\
\\
$C_{mn}$ & $mn/N$ \;\;\;\;\; & 2 \\
\hline
\end{tabular}
\caption{
\label{qpContent}
Some topological quantum numbers for quasiparticle excitations based
on considerations of Section \ref{Z2CS}.
$A_l$, for $l = 0, \cdots, 2N-1$, labels the $2N$ Abelian
quasiparticles. $B_l$, for $l = 0, \cdots, 2N-1$, labels the
$Z_2$ vortices. $C_{mn}$, for $m,n = 0, \cdots, N-1$ and $m < n$,
labels the $N(N-1)/2$ quasiparticles with quantum dimension $2$. 
Note that the quasiparticles $(e,m)$ in the Abelian $Z_N$ states
have spin $em/N$. Also note that the spin is meaningful only modulo 1. 
}
\end{table}

We still have not been able to compute the spins of the $Z_2$ vortices or
the complete fusion rules of the quasiparticles. In the following section,
we will present a prescription that enables us to calculate all of the 
topological properties of these twisted $Z_N$ states.

\section{Conformal field theory construction at $c - \bar{c} = 0$ }
\label{cftCon}

The use of CFT techniques to compute topological quantum numbers for FQH states
has been very powerful.\cite{MR9162,WW9455,LW1024} Physically, this is
possible because the edge theory is described by CFT, and there is a
correspondence between the spectrum of states in CFT and the topological
properties of quasiparticles in the bulk of FQH states.\cite{WWH9476} The
prescription in those cases is to identify an appropriate set of CFTs, choose
an appropriate electron operator, and then the quasiparticles are those
operators that can be constructed that are mutually local with respect to the
electron operator. Two quasiparticles that are related by electron operators
are topologically equivalent. The topological spin of the quasiparticles then
is believed to follow from the scaling dimension of the quasiparticle operator
in the CFT, while the fusion rules of the quasiparticles are equivalent to the
fusion rules, with respect to the electron chiral algebra, of the quasiparticle
operators in the CFT. 

In the case of the twisted $Z_N$ states, we do not expect to have topologically
protected edge modes.  However, under certain symmetry, gapless edge modes
described by CFT can exist.\cite{KL0834}  So the CFT prescription can still be
used to yield possible full sets of topological quantum numbers.  Physically,
we can think of this as the CFT that describes gapless edge excitations for
these states, although it is unstable to opening up a gap.  In this section, we
will give a prescription to compute the topological properties from CFT. While
we cannot prove that the topological quantum numbers are precisely those of the
$U(1) \times U(1) \rtimes Z_2$ CS theory, they are consistent with all of the
highly non-trivial results of the previous section. Additionally, based on the
relation of these twisted $Z_N$ states to their FQH counterparts, the orbifold
FQH states,\cite{BW1070} we have even more reason to believe that the
prescription given here is correct one. We expect it possible to prove that the
topological quantum numbers found using this prescription are in fact the
unique consistent set that are also consistent with results that can be deduced
from the $U(1) \times U(1) \rtimes Z_2$ CS theory. 

The construction is analogous to the orbifold FQH states,\cite{BW1070} except we
take the anti-holomorphic part of the $Z_2$ orbifold as the non-Abelian part of
the CFT instead of the holomorphic part; the ``charge'' part is the $c = 1$
chiral (holomorphic) scalar field. Thus the total central charge of the CFT is
$c_{tot} = c + \bar{c} = 2$, while the difference in central charges is
$c_{rel} = c - \bar{c} = 0$; this indicates that such a phase would have 0
thermal Hall conductance, as expected from the fact that it does not have
protected edge modes (see Section \ref{prEdgeSect}). 

We could also take the holomorphic part of the $Z_2$ orbifold
as the non-Abelian part, and the ``charge'' part to be anti-holomorphic. This
would yield the time-reversed counterpart of this phase. 

The operator content of the $Z_2$ orbifold CFT is reviewed in 
Appendix \ref{orbAppendix}. For the twisted $Z_N$ states, we take the ``electron'' operator to be:
\begin{align}
V_e(z, \bar z) = \bar{\phi}_N^1 (\bar z) e^{i \sqrt{\nu^{-1}} \varphi(z)},
\end{align}
where $\nu = 2/N$. The quasiparticle operators $V_q$ are those operators that are mutually
local with respect to the electron operator:
\begin{align}
V_q(z, \bar z) = \mathcal{O}(\bar z) e^{i Q \sqrt{\nu^{-1}} \varphi(z)}. 
\end{align}
The OPE of $V_q$ with $V_e$ is:
\begin{align}
V_q(w,\bar w) V_e(z, \bar z) \sim (w-z)^{Q/\nu} (\bar w - \bar z)^{h_{\mathcal{O}_2} - h_{\mathcal{O}} - h_{\bar{\phi}_N^1} } \mathcal{O}_2 + \cdots.
\end{align}
Thus for $V_q$ to be local w.r.t to $V_e$, we require:
\begin{align}
Q/\nu - (h_{\mathcal{O}_2} - h_{\mathcal{O}} - h_{\bar{\phi}_N^1}) = \text{ integer}.
\end{align}
Two quasiparticle operators are topologically equivalent if they can be related by the 
electron operator. Proceeding in this fashion, we find topological orders that
agree with the results of the previous section. This construction allows us to obtain all 
of the topological information of the twisted $Z_N$ phases. In the next section we list
examples of results that we obtain from this construction. 

We note that this is an interesting non-trivial example of the CS/CFT correspondence\cite{W8951}
because the boundary CFT in this case contains both holomorphic and anti-holomorphic 
parts that are glued together in a special way.

\section{Examples}

In this section, we list results obtained from the CFT consideration
for different twisted $Z_N$ states.

For $N=3$, the results are summarized in Table \ref{N3fields}.  We see that
there are 15 types of quasiparticles.  Those particles carry fractional angular
momentum which we call spin.\cite{Wtoprev} Note that the spin (or angular
momentum) does not have to be multiples of $\hbar/2$ in 2+1D.  The spin of a
quasiparticle can be measured by putting the system on a sphere or on other curved
spaces.

\begin{table}
\begin{center}
\begin{tabular}{llcl}
\hline
 & CFT Label & Quantum dim. & Spin \\
\hline
$\v 0$ & $\mathbb{I}$ &				1		& $0$ \\
$\v 1$ & $e^{i 2/3 \sqrt{3/2} \varphi}$ &	1		& $0 + 1/3 = 1/3$ \\
$\v 2$ & $\phi_N^2e^{i 1/3 \sqrt{3/2}\varphi}$ &	1		& $-3/4 + 1/12 \sim 1/3$ \\
$\v 3$ & $j$ &				1		& $-1 + 0 \sim 0$ \\
$\v 4$ & $je^{i 2/3 \sqrt{3/2}\varphi}$ &		1	& $-1 + 1/3 \sim 1/3$ \\
$\v 5$ & $\phi_N^1e^{i 1/3 \sqrt{3/2}\varphi}$ &		1	& $-3/4 +1/12 \sim 1/3$ \\
\\
$\v 6$ & $\sigma_1e^{i 1/2 \sqrt{3/2}\varphi}$ &	$\sqrt{3}$		& $-1/16 + 3/16 = 1/8$ \\
$\v 7$ & $\sigma_2e^{i 1/6 \sqrt{3/2}\varphi}$ &	$\sqrt{3}$		& $-1/16 + 1/48 = -1/24$ \\
$\v 8$ & $\sigma_2e^{i 5/6 \sqrt{3/2}\varphi}$ &	$\sqrt{3}$		& $-1/16 + 25/48 = 11/24$ \\
$\v 9$ & $\tau_1e^{i 1/2 \sqrt{3/2}\varphi}$ &	$\sqrt{3}$		& $-9/16 + 3/16 \sim 5/8$ \\
$\v 10$ & $\tau_2e^{i 1/6 \sqrt{3/2}\varphi}$ &	$\sqrt{3}$		& $-9/16 + 1/48 \sim 11/24$ \\
$\v 11$ & $\tau_2e^{i 5/6 \sqrt{3/2}\varphi}$ &	$\sqrt{3}$		& $-9/16 + 25/48 = -1/24$ \\
\\
$\v 12$ & $\phi_1e^{i 1/3 \sqrt{3/2}\varphi}$ &	2		& $-1/12 + 1/12 = 0$ \\
$\v 13$ & $\phi_2e^{i 0 \sqrt{3/2}\varphi}$ &	2		& $-1/3 + 0 \sim 2/3$ \\
$\v 14$ & $\phi_2e^{i 2/3 \sqrt{3/2}\varphi}$ &		2	& $1/3 - 1/3 = 0$ \\
\hline
\end{tabular}
\caption{
\label{N3fields}
Quasiparticle operators for CFT construction of twisted $Z_3$ phase. 
}
\end{center}
\end{table}

For $N=2$, we have $9$ quasiparticles, as summarized in Table \ref{N2fields}. 
\begin{table}
\begin{center}
\begin{tabular}{llccc}
\hline
 & CFT Label & Q. dim. & Spin & $Ising \times \overline{Ising}$ fields \\
\hline
$\v 0$ & $\mathbb{I}$ &		1				& $0 + 0 = 0$ & $\mathbb{I} \otimes \mathbb{I}$\\
$\v 1$ & $\phi_N^1$ &			1			& $-1/2 + 0 = 1/2$  & $\psi \otimes \mathbb{I}$\\
$\v 2$ & $j$ &			1			& $-1 + 0 \sim 0$ & $\psi \otimes \bar \psi$\\
$\v 3$ & $\phi_N^2$ &				1		& $-1/2 + 0 \sim 1/2$ & $\mathbb{I} \otimes \bar \psi$\\
\\
$\v 4$ & $\sigma_1e^{i 1/2 \sqrt{2} \varphi}$ &	$\sqrt{2}$		& $-1/16 +1/8 = 1/16$ & $\sigma \otimes \mathbb{I}$\\ 
$\v 5$ & $\sigma_2$ &			$\sqrt{2}$			& $-1/16 + 0 = -1/16$ & $\sigma \otimes \bar \psi $\\
$\v 6$ & $\tau_2$ &			$\sqrt{2}$			& $-9/16 + 0 = -9/16$ & $\mathbb{I} \otimes \bar \sigma$\\
$\v 7$ & $\tau_1e^{i 1/2 \sqrt{2}\varphi}$ &	$\sqrt{2}$     & $-9/16 + 1/8  \sim 9/16$  & $\psi \otimes \bar \sigma$\\
\\
$\v 8$ & $\phi_1e^{i 1/2 \sqrt{2}\varphi}$ &      2		& $-1/8 + 1/8 = 0$ & $\sigma \otimes \bar \sigma$\\
\hline
\end{tabular}
\caption{
\label{N2fields}
Quasiparticle operators for CFT construction of twisted $Z_2$ phase. Note
this is equivalent to $Ising \times \overline Ising$.  
}
\end{center}
\end{table}
It appears that this coincides with the $Ising \times \overline{Ising}$ topological order. Condensation
of the boson $\psi \otimes \bar \psi = j$ yields the $Z_2$ topological order.

\section{Discussion}
\label{Disc}

\subsection{Transition to twisted $Z_N$ topological phases }

Let $\gamma$ denote an anyon with statistical angle $\theta = 2\pi/N$ in
a topological phase, and let
$m$ control the mass of, or energy gap to creating, $\gamma$. As we
tune $m$, $\gamma$ may condense and drive a phase transition to a new phase. 
This transition can be described by the 
$\langle \phi \rangle = 0 \rightarrow \langle \phi \rangle \neq 0$ 
transition in a Chern-Simons Ginzburg-Landau theory:
\begin{align}
\label{sCS} 
\cL &=  |(\prt_0 + i a_0 )\phi|^2 -  v^2 |(\prt_i + i a_i)\phi|^2 - f |\phi|^2 - g |\phi|^4 
\nonumber\\ 
&  -\frac{\pi}{\th} \frac{1}{4\pi} a_\mu\prt_\nu a_\la \eps^{\mu\nu\la}.
\end{align}
In the above Lagrangian, the anyon number is conserved
In this case, the anyon condensation induces a transition between
Abelian states described by different $K$-matrices. 

In the case where $\gamma$ is only conserved modulo $N$, there will be an
additional term in the Lagrangian:
\begin{align}
\del \cL= t (\phi \hat M)^{N}+h.c.
\end{align}
In this case, the anyon condensation may induce a transition between
Abelian and non-Abelian states. 

In our study of bilayer quantum Hall phase transitions in \Ref{BW1004}, it was
suggested that this transition, in the presence of the $\delta \mathcal{L}$
term, may be dual to a 3D Ising transition. In those cases, one starts from an
Abelian bilayer FQH phase and obtains the non-Abelian orbifold FQH states by
tuning the interlayer tunneling and/or interlayer repulsion.  We may obtain a
similar situation in the context of $Z_N$ gauge theory if we reduce the energy
gap to the $(1,1)$ quasiparticles (the bound state of a single electric and a
single magnetic quasiparticle). The $(1,1)$ quasiparticles are conserved only
modulo $N$, because there is no additional conserved $U(1)$ charge as in the
FQH phases. This implies the possibility of an analog of the bilayer $(NN0)$
FQH phase transitions studied earlier but for a system in the absence of a
magnetic field and with no protected edge modes. The $(1,1)$ quasiparticle
plays the role of the f-exciton, both of which have statistical angle $\theta =
2\pi/N$. Tuning the interlayer repulsion is equivalent to tuning the attraction
between the minimal electric and magnetic quasiparticles. 

Note that while the $Z_N$ phase can be obtained in a time-reversal invariant
system, condensing the $(1,1)$ quasiparticle breaks time-reversal for $N > 2$. 

Therefore, consider starting with the Hamiltonian that gives deconfined $Z_N$,
and adding a term that can tune the attraction between the minimal electric and
magnetic quasiparticles. This will reduce the energy gap to their bound state,
and may be used to tune through a 3D Ising phase transition. The phase that
appears after the transition, in analogy to the bilayer FQH cases, may be the
twisted $Z_N$ gauge theory, described by $U(1) \times U(1) \rtimes Z_2$
Chern-Simons theory.  

\subsection{Time-reversal invariance}

We see that for $N > 2$, the topological quantum numbers break time reversal
symmetry -- there is no way that a topological phase with these quantum numbers
can preserve time-reversal symmetry.  In fact, we saw that we had a choice of
whether to pick the holomorphic part of the $Z_2$ orbifold and the
anti-holomorphic part of the $U(1)$ sector, or vice versa. This fact at first
appears worrisome, because these phases are separated from the $Z_N$ Abelian
phases through a 3D Ising transition, and the $Z_N$ phases are time-reversal
invariant phases. In the following we outline reasons to believe that indeed
these twisted $Z_N$ phases are not time-reversal invariant for $N > 2$. 

First, observe that for $N > 2$, the number of quasiparticles in these phases
is not a perfect square.  Typically, almost all time-reversal invariant
topological phases are ``doubled'' theories in the sense that mathematically
they are described by $G \otimes \bar{G}$ modular tensor categories, where $G$
is itself a modular tensor category and $\bar{G}$ is its time-reversed partner.
More in depth considerations also suggest that for $N > 2$, there is no
consistent topological phase that is time-reversal invariant and that has
$N(N+7)/2$ quasiparticles with the quantum dimensions described in Section
\ref{Z2CS}.\cite{mLevin} 

In addition to general considerations of what mathematically consistent
time-reversal invariant topological phases can exist, also note that the only
way that we currently know how to describe the $U(1) \times U(1) \rtimes Z_2$
CS theory from a microscopic starting point is through a slave-Ising/parton
construction, where partons are put into $\nu = \pm 1$ IQH states. Such a
$UV$-completion necessarily breaks time-reversal symmetry, so it is consistent
to find phases that cannot exist in the presence of time-reversal symmetry.  In
the case of the $Z_N$ Abelian phase, there are other microscopic realizations
of such topological order that do preserve time-reversal symmetry. 

Finally, note that the picture that we developed for the transition from the
$Z_N$ phase to the twisted $Z_N$ phase involved the condensation of a
particular anyon that has spin $1/N$. Thus for $N > 2$, putting this anyon into
some collective state will necessarily break time-reversal symmetry,
unless the anyon with spin $-1/N$ is treated on exactly the same footing. 

\subsection{Protected edge modes} \label{prEdgeSect}

The $Z_N$ Abelian phase does not have protected gapless edge modes in the
absence of any symmetries, and here we have seen that it is separated from the
twisted $Z_N$ non-Abelian phases by a $Z_2$ transition.  Viewed from the
twisted phase, the transition can be thought of as the condensation of a boson
$j$ that squares to a topologically trivial excitation. On general grounds, we
expect that the boundary between two topological phases will not have protected
gapless edge modes if the two phases are related by a $Z_N$ boson condensation
transition.\cite{mLevin} Since the $Z_N$ phase does not have protected gapless
edge modes at a boundary with the vacuum, this means that the twisted $Z_N$
phase will also not have protected gapless edge modes at a boundary with the
vacuum. 

We expect that the above discussion can be made more concrete by studying the
edge through the $U(1) \times U(1) \rtimes Z_2$ CS theory and the slave-Ising
theory and showing that all possible gapless edge modes can be gapped out by
allowed perturbations.

\section{Summary, Conclusion, and Outlook}

We have seen that the deconfined phase of $Z_N$ gauge theories has a 
neighboring non-Abelian phase, the twisted $Z_N$ states. These
two phases are separated by a continuous quantum phase transition
and the non-Abelian states can be accessed, for $N > 2$, only
by breaking time-reversal symmetry. 

In this paper, we have studied the full topological order of these
non-Abelian states. Much of the topological order can be deduced
directly from the $U(1) \times U(1) \rtimes Z_2$ CS theory and
the fact that it is separated from the conventional $Z_N$ 
states by the condensation of a $Z_2$-charged boson. We found
a way to compute the rest of the topological properties that we could
not calculate directly, although those results rely on additional
assumptions.

In addition to deriving the topological order of these states,
we presented a parton construction that allows us to describe 
the $Z_N$ topological order in terms of fermions in band
insulators with Chern number $\pm 1$. This description of the
$Z_N$ states then allowed us to describe the non-Abelian twisted $Z_N$
through a slave Ising theory of $Z_2$ fractionalization. Such a 
construction provides trial projected wave functions and helps 
establish that these phases are physical in that they can be
realized in bosonic systems with local interactions. 

There are two main conceptual issues lacking in our understanding
of these states. First, we should be able to prove more rigorously 
that the full topological quantum numbers presented here coincide
with those of the $U(1) \times U(1) \rtimes Z_2$ CS theory and
the associated slave Ising description. Second, and more importantly,
we would like to understand better how to access these non-Abelian 
states by starting from the Abelian $Z_N$ states. We know little besides
the fact that the energy gap of the $(1,1)$ quasiparticles should probably
be tuned through zero. 

In the case of the $Z_N$ topological order, we found a way through field
theoretic and slave-particle constructions to essentially gauge
the electric-magnetic symmetry of the topological quantum numbers. 
However, conceptually we do not know how to extend these ideas to
other discrete gauge theories. It would be interesting to develop
more general theoretical, physical descriptions that allows us to ``twist''
the symmetries of the topological quantum numbers of a phase.
In CFT, such a procedure is referred to as orbifolding. In the context
of bulk $2+1$-dimensional states of matter, we do not have any physical
understanding of how this can be done more generally. One starting
point would obviously be to try to develop Chern-Simons descriptions
of discrete gauge theories, in the way that the mutual $U(1) \times U(1)$
CS theory describes $Z_N$ gauge theory. 

Recently, another series of topological phase transitions was found involving the non-abelian $SU(2)_N \times \overline{SU(2)}_N$
states, where the transition involves the condensation of a boson with $Z_2$ fusion rules.\cite{BS1017} 
By explicitly constructing a lattice model, it was found that the condensation of the boson
yields a continuous phase transition in the 3D Ising universality class. For $N = 2$, the
results of \Ref{BS1017} coincide with our results. However the generalization to $N > 2$ is different;
in our case, the $N > 2$ twisted $Z_N$ states break time-reversal symmetry though they have no topologically 
protected edge modes, and the states on the other side of the transition are described by $Z_N$ gauge theory.
On the other hand, the $SU(2)_N \times \overline{SU(2)}_N$ states can always exist in time-reversal invariant systems,
and the states on the other side of the phase transition are not describable by $Z_N$ gauge theory for $N > 2$.

We thank Michael Levin for helpful discussions. 
XGW is supported by  NSF Grant No.
DMR-1005541. MB is supported by a fellowship from the Simons
Foundation. 

\appendix

\section{Operator content of $U(1)/Z_2$ orbifold CFT}
\label{orbAppendix}

Since the $U(1)/Z_2$ orbifold at $c = 1$ plays an important role in
understanding the topological properties of the twisted $Z_N$ states, 
here we will give a brief account of some of its properties. 
The information here is taken from \Ref{DV8985},
where a more complete discussion can be found. 

The $U(1)/Z_2$ orbifold CFT, at central charge $c = 1$, is the theory 
of a scalar boson $\varphi$, compactified at a radius $R$, so that 
$\varphi \sim \varphi + 2\pi R$, and with an additional $Z_2$ gauge 
symmetry: $\varphi \sim -\varphi$. When $\frac{1}{2} R^2$ is 
rational, \it i.e. \rm $\frac{1}{2} R^2 = p/p'$, 
with $p$ and $p'$ coprime, then it is useful to
consider an algebra generated by the fields $j = i \partial \varphi$,
and $e^{\pm i \sqrt{2N} \varphi}$, for $N = p p'$. This algebra is referred to
as an extended chiral algebra. The infinite number of Virasoro primary fields
in the $U(1)$ CFT can now be organized into a finite number of representations 
of this extended algebra $\mathcal{A}_N$. There are $2N$ of these 
representations, and the primary fields are written as $V_k = e^{ik\varphi/\sqrt{2N}}$, 
with $k = 0, 1, \cdots, 2N-1$. The $Z_2$ action takes $V_k \rightarrow V_{2N-k}$.

In the $Z_2$ orbifold, one now considers representations of the smaller
algebra $\mathcal{A}_N/Z_2$. This includes the $Z_2$ invariant combinations of the
original primary fields, which are of the form $\phi_k = \cos(k \varphi/\sqrt{2N})$;
there are $N+1$ of these. In addition, there are 6 new primary fields. The gauging
of the $Z_2$ allows for twist operators that are not local with respect to the fields
in the algebra $\mathcal{A}_N/Z_2$, but rather local up to an element of $Z_2$. It turns
out that there are two of these twisted sectors, and each sector contains one field
that lies in the trivial representation  of the $Z_2$, and one field that lies
in the non-trivial representation of $Z_2$. These twist fields are labelled
$\sigma_1$, $\tau_1$, $\sigma_2$, and $\tau_2$. In addition to these, an in-depth
analysis \citep{DV8985} shows that the fixed points of the $Z_2$ action in the original $U(1)$ 
theory split into a $Z_2$ invariant and a non-invariant field. We have already 
counted the invariant ones in our $N+1$ invariant fields, which leaves 2 new
fields. One fixed point is the identity sector, corresponding to $V_0$, which 
splits into two sectors: $1$, and $j = i \partial \varphi$. The other
fixed point corresponds to $V_N$. This splits into two primary fields, which 
are labelled as $\phi_N^i$ for $i = 1,2$ and which have scaling dimension $N/4$.
In total, there are $N+7$ primary fields in the $Z_2$ rational orbifold at ``level'' $2N$.
These fields and their properties are summarized in Table \ref{Z2orbFields}.
\begin{table}
\begin{center}
\begin{tabular}{ccc}
\hline
Label & Scaling Dimension & Quantum Dimension \\
\hline
$\mathbb{I}$ & 0 & 1 \\
$j$ & 1 & 1 \\
$\phi_N^1$ & $N/4$ & 1 \\
$\phi_N^2$ & $N/4$ & 1 \\
\\
$\sigma_1$ & 1/16 & $\sqrt{N}$ \\
$\sigma_2$ & 1/16 & $\sqrt{N}$ \\
$\tau_1$ & 9/16 & $\sqrt{N}$ \\
$\tau_2$ & 9/16 & $\sqrt{N}$ \\
\\
$\phi_k$ & $k^2/4N$ & $2$ \\
\hline
\end{tabular}
\end{center}
\caption{\label{Z2orbFields}
Primary fields in the $U(1)_{2N}/Z_2$ orbifold CFT. The label $k$ runs
from $1$ to $N-1$.}
\end{table}

This spectrum for the $Z_2$ orbifold is obtained by first computing the
partition function of the full $Z_2$ orbifold CFT defined on a torus, 
including both holomorphic and anti-holomorphic parts. Then, the partition
function is decomposed into holomorphic blocks, which are conjectured to
be the generalized characters of the $\mathcal{A}_N/Z_2$ chiral algebra. 
This leads to the spectrum listed in Table \ref{Z2orbFields}. 
The fusion rules and scaling dimensions for these primary fields are obtained 
by studying the modular transformation properties of the characters.

The fusion rules are as follows. For $N$ even:
\begin{align}
j \times j  &=1,
\nonumber \\
\phi^i_N \times \phi^i_N &= 1,
\nonumber \\
\phi^1_N \times \phi_N^2 &= j.
\end{align}
As mentioned in \Ref{DV8985}, the vertex operators $\phi_k$ have a fusion algebra
consistent with their interpretation as $\cos \frac{k}{\sqrt{2N}} \varphi$,
\begin{align}
\phi_k \times \phi_{k'} &= \phi_{k+k'} + \phi_{k - k'} \;\;\;\;\; (k' \neq k, N- k),
\nonumber \\
\phi_k \times \phi_k &= 1 + j + \phi_{2k},
\nonumber \\
\phi_{N-k} \times \phi_k &= \phi_{2k} + \phi^1_N + \phi^2_N,
\nonumber \\
j \times \phi_k &= \phi_k.
\end{align}
\begin{align}
\sigma_i \times \sigma_i &= 1 + \phi_N^i + \sum_{k \text{ even}} \phi_k,
\nonumber \\
\sigma_1 \times \sigma_2 &= \sum_{k \text{ odd}} \phi_k,
\nonumber \\
j \times \sigma_i &= \tau_i
\end{align}

\begin{table}[t]
\begin{tabular}{ccc}
\hline
$Z_2$ Orb. field & Scaling Dimension, $h$ & $Z_4$ parafermion field \\
\hline
\hline
$1$ & $0$ & $\Phi^0_0$ \\
$j$ & $1$ & $\Phi^0_4$ \\
$\phi^1_N$ & $3/4$ & $\Phi^0_2$ \\
$\phi^2_N$ & $3/4$ & $\Phi^0_6$ \\
$\phi_1$ & $1/12$ & $\Phi^2_2$ \\
$\phi_2$ & $1/3$ & $\Phi^2_0$ \\
$\sigma_1$ & $1/16$ & $\Phi^1_1$ \\
$\sigma_2$ & $1/16$ & $\Phi^1_{-1}$ \\
$\tau_1$ & $9/16$ & $\Phi^1_3$ \\
$\tau_2$ & $9/16$ & $\Phi^1_5$ \\
\hline
\end{tabular}
\caption{
\label{Z2fieldsN3}
Primary fields in the $Z_2$ orbifold for $N=3$, their 
scaling dimensions, and the $Z_4$ parafermion fields that they correspond to.}
\end{table}

For $N$ odd, the fusion algebra of 1, $j$, and $\phi_N^i$ is
$Z_4$:
\begin{align}
j \times j &= 1,
\nonumber \\
\phi_N^1 \times \phi_N^2 &= 1,
\nonumber \\
\phi_N^i \times \phi_N^i &= j.
\end{align}
The fusion rules for the twist fields become:
\begin{align}
\sigma_i \times \sigma_i &= \phi_N^i + \sum_{k \text{ odd}} \phi_k,
\nonumber \\
\sigma_1 \times \sigma_2 &= 1 + \sum_{k \text{ even}} \phi_k.
\end{align}
The fusion rules for the operators $\phi_k$ are unchanged. 

For $N = 1$, it was observed that the $Z_2$ orbifold is equivalent
to the $U(1)_8$ Gaussian theory. For $N = 2$, it was observed 
that the $Z_2$ orbifold is equivalent to two copies of the 
Ising CFT. For $N = 3$, it was observed that the $Z_2$ orbifold 
is equivalent to the $Z_4$ parafermion CFT of Zamolodchikov
and Fateev.\citep{ZF8515}

In Tables \ref{Z2fieldsN2} and \ref{Z2fieldsN3} we list the
fields from the $Z_2$ orbifold for $N=2$ and $N=3$, their
scaling dimensions, and the fields in the $Ising^2$ or $Z_4$ 
parafermion CFTs that they correspond to.

\begin{table}
\begin{tabular}{ccc}
\hline
$Z_2$ Orb. field & Scaling Dimension, $h$ & $Ising^2$ fields \\
\hline
\hline
$1$ & $0$ & $\mathbb{I} \otimes \mathbb{I}$ \\
$j$ & $1$ & $\psi \otimes \psi$ \\
$\phi^1_N$ & $1/2$ & $\mathbb{I} \otimes \psi$ \\
$\phi^2_N$ & $1/2$ & $\psi \otimes \mathbb{I}$ \\
$\phi_1$ & $1/8$ & $\sigma \otimes \sigma$ \\
$\sigma_1$ & $1/16$ & $\sigma \times \mathbb{I}$ \\
$\sigma_2$ & $1/16$ & $\mathbb{I} \otimes \sigma$ \\
$\tau_1$ & $9/16$ & $\sigma \otimes \psi$ \\
$\tau_2$ & $9/16$ & $\psi \otimes \sigma$ \\
\hline
\end{tabular}
\caption{
\label{Z2fieldsN2}
Primary fields in the $Z_2$ orbifold for $N=2$, their 
scaling dimensions, and the fields from Ising$^2$ to which they
correspond.}
\end{table}

\section{Ground state degeneracy on a torus for $SU(N)^3 \times U(1)$ gauge theory}
\label{NAtorusCalc}

A procedure for calculating the ground state degeneracy on a torus
for states obtained through the projective construction was described in \Ref{W9927}. 
This procedure works for gauge groups that are connected, while
gauge groups of the form $G \rtimes H$, where $G$ is connected and 
$H$ is a discrete group, require further analysis.

The classical configuration space of CS theory consists of flat connections,
for which the magnetic field vanishes: $\epsilon_{ij} \partial_i a_j = 0$. 
This configuration space is completely characterized by holonomies of the
gauge field along the non-contractible loops of the torus:
\begin{align}
W(\alpha) = \mathcal{P} e^{i \oint_\alpha a \cdot dl}.
\end{align}
More generally, for a manifold $M$, the gauge-inequivalent set of $W(\alpha)$ form
a group: $(\text{Hom: } \pi_1(M) \rightarrow G) /G$, which is the group of
homomorphisms of the fundamental group of $M$ to the gauge group $G$,
modulo $G$. For a torus, $\pi_1(T^2)$ is Abelian, which means that 
$W(\alpha)$ and $W(\beta)$, where $\alpha$ and $\beta$ are the two distinct
non-contractible loops of the torus, commute with each other and we can always 
perform a global gauge transformation so that $W(\alpha)$ and $W(\beta)$
lie in the maximal Abelian subgroup, $G_{abl}$, of $G$ (this subgroup is called the maximal torus).  
The maximal torus is generated by the Cartan subalgebra of the Lie algebra of $G$; in the case at
hand, this Cartan subalgebra is composed of $3N-2$ matrices, $3(N-1)$ of
which lie in the Cartan subalgebra of $SU(N) \times SU(N) \times SU(N)$, 
in addition to $diag(1, 0,..., 1, 0,...,-1,0,...)$. Since we only need to consider
components of the gauge field $a^I$ that lie in the Cartan subalgebra,
the CS Lagrangian becomes 
\begin{align} \mathcal{L} = \frac{1}{4\pi} K_{IJ} a^I \partial a^J,
\end{align}
where $K_{IJ} = \text{Tr}(p^I p^J)$ and $p^I$, $I = 1, \cdots, k+1$ are the 
generators that lie in the Cartan subalgebra. 

There are large gauge transformations $U =  e^{2\pi x_i p^I/L}$, where $x_1$
and $x_2$ are the two coordinates on the torus and $L$ is the length of each
side. These act on the partons as
\begin{equation}
\psi \rightarrow U \psi,
\end{equation}
where $\psi^T = (\psi_1, \cdots, \psi_{3N})$, and they take 
$a^I_i \rightarrow a^I_i + 2\pi/L$. These transformations will be the
minimal large gauge transformations if we normalize the generators as follows:
\begin{align}
p^I_{ij} &= \delta_{ij} (\delta_{i,I} - \delta_{i, I + 1}), \;\;\;\;\; I = 1, \cdots, N-1,
\nonumber \\
p^I_{ij} &= \delta_{ij} (\delta_{i,I+1} - \delta_{i,I+2}), \;\;\;\;\; I = N, \cdots, 2N-2,
\nonumber \\
p^I_{ij} &= \delta_{ij} (\delta_{i,I+2} - \delta_{i,I+3}), \;\;\;\;\; I = 2N-1, \cdots, 3N-3,
\nonumber \\
p^{3N - 2}_{ij} &= \delta_{ij} (\delta_{i,1} + \delta_{i,N+1} - \delta_{i, 2N +1})
\end{align}

The effective $K$-matrix is of the form
\begin{align}
\label{KmatrixZN}
K = \left( \begin{matrix} 
A & 0 & 0 & v \\ 
0 & A & 0 & v \\
0 & 0 & -A & v \\
v^T & v^T & v^T & 1 \\
\end{matrix} \right),
\end{align}
where $A$ is the Cartan matrix of $SU(N)$ (an $N-1 \times N-1$ matrix), 
and $v$ is an $(N-1) \times 1$ column vector with 1 on the first 
entry and 0s everywhere else: $v^T = (1, 0, ..., 0)$. For example, 
for $N = 2$ the above $K$-matrix is
\begin{align}
\left( \begin{matrix}
2 & 0 & 0 & 1 \\
0 & 2 & 0 & 1 \\
0 & 0 & -2 & 1 \\
1 & 1 & 1 & 1 \\
\end{matrix} \right).
\end{align}
For $N = 4$, it is
\begin{align}
\left( \begin{matrix}
2  & -1 & 0  & 0 & 0 & 0 & 0 & 0 & 0 & 1 \\ 
-1 & 2  & -1 & 0 & 0 & 0 & 0 & 0 & 0 & 0 \\
0 & -1  &  2 & 0 & 0 & 0 & 0 & 0 & 0 & 0 \\
0 &    0  &   0  &   2 &   -1  &   0  &   0  &   0  &   0  &   1 \\
0 &    0  &   0  &  -1 &    2  &  -1  &   0  &   0  &   0  &   0 \\
0 &    0  &   0  &   0 &   -1  &   2  &   0  &   0  &   0  &   0 \\
0 &    0  &   0  &   0 &    0  &   0  &  -2  &   1  &   0  &   1 \\
0 &    0  &   0  &   0 &    0  &   0  &   1  &  -2  &   1  &   0 \\
0 &    0  &   0  &   0 &    0  &   0  &   0  &   1  &  -2  &   0 \\
1 &    0  &   0  &   1 &    0  &   0  &   1  &   0  &   0  &   1 \\
\end{matrix} \right).
\end{align}

In addition to the large gauge transformations, there are discrete gauge
transformations $W \in SU(N) \times SU(N) \times SU(N) \times U(1)$
which keep the Abelian subgroup unchanged but interchange the $a^I$'s amongst themselves. 
These satisfy
\begin{align}
W^{\dagger} G_{abl} W = G_{abl},
\end{align}
or, alternatively,
\begin{align}
\label{dscTrZn}
W^{\dagger} p^I W = T_{IJ} p^J,
\end{align}
for some $(3N-2) \times (3N-2)$ matrix $T$. These discrete transformations 
correspond to the independent ways of interchanging the partons and they
correspond to the Weyl group of the gauge group. The Weyl group for 
$SU(N)$ is $S_N$. These can be generated by pairwise interchanges of
the partons. 

Picking the gauge $a_0^I = 0$ and parametrizing the gauge field as 
\begin{align}   
a^I_1 = \frac{2\pi}{L}X^I_1 \;\;\;\;\; a^I_2 = \frac{2\pi}{L} X^I_2, 
\end{align}
we have
\begin{align}
L = 2 \pi K_{IJ} X^I_1 \dot X^J_2.
\end{align}
The Hamiltonian vanishes. The conjugate momentum to $X^J_2$ is
\begin{align}
p^J_{2} = 2 \pi K_{IJ} X^I_1.
\end{align}
Since $X_2^J \sim X_2^J + 1$ as a result of the large gauge transformations, 
we can write the wave functions as
\begin{align}
\psi( \vec X_2) = \sum_{\vec n} c_{\vec n} e^{2\pi \vec n \cdot \vec X_2},
\end{align}
where $\vec X_2 = (X_2^1, \cdots X_2^{2N-3})$ and $\vec n$ is a $(2N-3)$-dimensional vector of integers. 
In momentum space the wave function is
\begin{align}
\phi( \vec p_2 ) &= \sum_{\vec n} c_{\vec n} \delta^{(2N-3)}( \vec p_{2} - 2 \pi \vec n)
\nonumber \\
 &\sim \sum_{\vec n} c_{\vec n} \delta^{(2N-3)}( K \vec X_1 - \vec n),
\end{align}
where $\delta^{(2N-3)}(\vec x)$ is a $(2N-3)$-dimensional delta function. 
Since $X_1^J \sim X_1^J + 1$, it follows that $c_{\vec n} = c_{\vec n'}$, where
$(\vec n')^I = n^I + K_{IJ}$, for any $J$. Furthermore, each discrete gauge transformation 
$W_i$ that keeps the Abelian subgroup $G_{abl}$ invariant corresponds to a 
matrix $T_i$ (see eqn. \ref{dscTrZn}), which acts on the diagonal generators.
These lead to the equivalences $c_{\vec n} = c_{T_i \vec n}$. 

Carrying out the result on the computer, we find that $\text{Det } K$ is always
equal to $N^2$, and, remarkably, we find that the Weyl group, \it ie \rm the 
group of discrete transformations that keeps the Abelian subgroup unchanged, 
acts trivially in the sense that it does not lead to any identifications 
among the states. This suggests that the $K$-matrix is a complete description 
of the theory on a torus!


%

\end{document}